\documentclass[sigconf, nonacm]{acmart}

\usepackage{subfiles}

\usepackage{caption}
\usepackage{subcaption}
\usepackage{multirow}
\usepackage{makecell}
\usepackage{tabularx}
\usepackage{csquotes}

\usepackage{xcolor}
\definecolor{lightergray}{RGB}{240, 240, 240}

\usepackage{listings}
\newcommand{\squote}[1]{`#1'}
\newcommand{\dquote}[1]{``#1''}

\newcommand{\pquote}[2]{``#1''~(P#2)}
\newcommand{\pcount}[1]{\n{#1}}

\newcommand{\sys}{\textit{ION}}
\newcommand{\hai}{\textsc{AI Highlight}}
\newcommand{\huser}{\textsc{User Highlight}}
\newcommand{\hchart}{\textsc{Summary Chart}}
\newcommand{\hsays}{\textsc{Fallacy Explanation}}
\newcommand{\psug}{\textsc{Suggested Queries}}
\newcommand{\pwrite}{\textsc{Write Own Query}}
\newcommand{\pweb}{\textsc{Web Findings}}
\newcommand{\cfood}{\textsc{Food For Thought}}
\newcommand{\cdis}{\textsc{Discussion Space}}

\newcommand{\none}{\textit{None}}
\newcommand{\nth}{\textit{Nothing}}
\newcommand{\DG}[2]{\textbf{DG#1: #2}}

\newcommand{\n}[1]{$n = #1$}
\newcommand{\N}[1]{$N = #1$}
\newcommand{\MSD}[2]{$M = #1, SD = #2$}
\newcommand{\MIQR}[2]{$med = #1, IQR = #2$}
\newcommand{\ttest}[2]{$t = #1, p #2$}
\newcommand{\Wilcox}[2]{$W = #1, p #2$}

\newcommand{\Pearson}[3]{$r(#1) = #2, p #3$}
\newcommand{\Spearman}[3]{$r_s(#1) = #2, p #3$}

\newcommand{\ncs}{\textit{Need for Cognition}}
\newcommand{\ed}{\textit{Ease of Discussing Topic}}
\newcommand{\ia}{\textit{Informational Awareness}}

\newcommand{\ix}{\textit{Interactions}}
\newcommand{\lik}{\textit{Likability}}
\newcommand{\use}{\textit{Usefulness}}

\AtBeginDocument{
  }

\begin{document}

\title{Iffy-Or-Not: Extending the Web to Support the Critical Evaluation of Fallacious Texts}

\author{Gionnieve Lim}
\email{gionnievelim@gmail.com}
\orcid{0000-0002-8399-1633}
\affiliation{
    \institution{Singapore University of Technology and Design}
    \country{Singapore}
}

\author{Juho Kim}
\email{juhokim@kaist.ac.kr}
\orcid{0000-0001-6348-4127}
\affiliation{
  \institution{School of Computing, KAIST}
  \city{Daejeon}
  \country{Korea}
}

\author{Simon T. Perrault}
\email{perrault.simon@gmail.com}
\orcid{0000-0002-3105-9350}
\affiliation{
    \institution{Singapore University of Technology and Design}
    \country{Singapore}
}

\renewcommand{\shortauthors}{Lim et al.}

\begin{abstract}
Social platforms have expanded opportunities for deliberation with the comments being used to inform one’s opinion. However, using such information to form opinions is challenged by unsubstantiated or false content. To enhance the quality of opinion formation and potentially confer resistance to misinformation, we developed Iffy-Or-Not (\sys{}), a browser extension that seeks to invoke critical thinking when reading texts. With three features guided by argumentation theory, \sys{} highlights fallacious content, suggests diverse queries to probe them with, and offers deeper questions to consider and chat with others about. From a user study (\N{18}), we found that \sys{} encourages users to be more attentive to the content, suggests queries that align with or are preferable to their own, and poses thought-provoking questions that expands their perspectives. However, some participants expressed aversion to \sys{} due to misalignments with their information goals and thinking predispositions. Potential backfiring effects with \sys{} are discussed.
\end{abstract}



\begin{teaserfigure}
  \includegraphics[width=\textwidth]{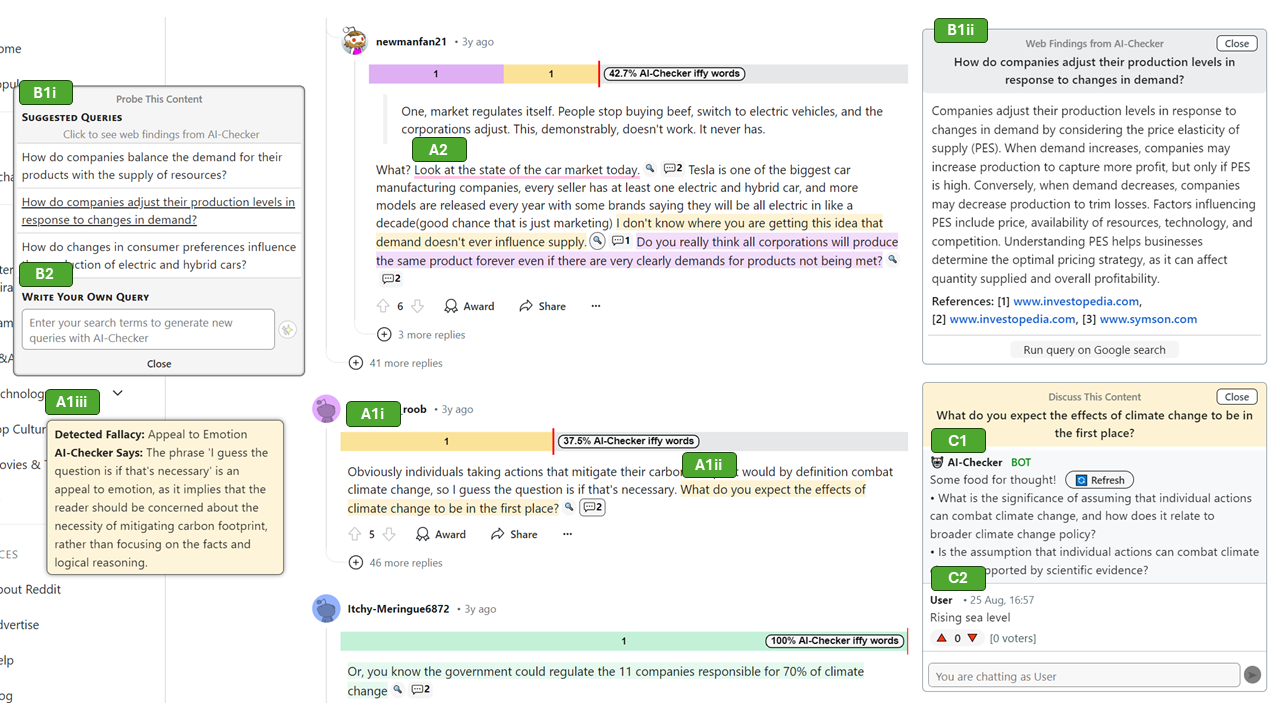}
  \caption{Iffy-Or-Not (\sys{}) is a browser extension with several features that promote critical thinking: (A) highlight, (B) probe, and (C) chat. When enabled, it scans the page. It then highlights content containing fallacies and (A1i) provides an overview of the number of fallacies detected. Users can then (A1ii) hover over the highlight to (A1iii) get more details about why the content is fallacious. Users can also (A2) author problematic content and provide a reason explaining their concerns. The magnifying glass button allows users to consider more contextual information through (B1i) queries suggested by \sys{} that are coupled with (B1ii) summaries drawn from the web. They may also (B2) enter their own search terms to get customized variations of the queries. The speech bubble button provides users with (C1) questions to challenge the content, which the user can (C2) respond to and vote on. \sys{} supports the user to think more critically about content to form their own opinions and veracity judgments of them.}
  \Description{Figure showing the features of Iffy-Or-Not in an online forum. Some parts of the comments have AI and User highlights. The Fallacy Explanation, Suggested Queries and Write Your Own Query are shown on the right of the comments. The Food-For-Thought and Discussion Space, and Web Findings are shown on the right of the comments.}
  \label{fig:system}
\end{teaserfigure}

\maketitle

\section{Introduction}

Social media, forums and comment sections on websites have become spaces for online deliberation with the extensive reach and ease of access to these platforms leading to the confluence of individuals with unique backgrounds and experiences who can share a great variety of insights and perspectives~\cite{Wright2007, Halpern2013, Semaan2014}. The wide and diverse range of opinions online makes it a valuable resource for people to understand societal issues from different points of view and use it to inform their own opinions on them~\cite{Surowiecki2004}.

While such platforms can act as spaces for deliberation, the quality of the information present can be undermined by various forms of problematic content such as spam, incendiary and demeaning content, and most crucially, misinformation. While the earlier examples may impede information processing as more effort is needed to filter them away, misinformation has direct impact on the opinions formed as beliefs that manifest from misinformation are unfounded~\cite{McKay2020}. Misinformation can influence attitudes toward societal issues with real consequences as observed with the increasing polarization of democratic societies globally~\cite{Carothers2019b}.

With misinformation being a pervasive issue on social platforms, several approaches have been set in place to combat it. This includes having volunteer moderators or engaging professionals to fact-check and label problematic content~\cite{Carson2023}, using algorithms to downrank and remove them~\cite{Vincent2022}, and having policies against abuses of the platforms or face consequences like bans~\cite{Vincent2021}. While there are many measures, they are limited in various ways such as the scalability of human effort being bounded due to the small number of people compared to the tremendous volume of information~\cite{Moy2021} and detection algorithms having to be sensitive to misinformation that changes with the times that may now even be further escalated by LLMs~\cite{Hombaiah2021, Chen2024}.
Misinformation continues to be prevalent and with no single measure being a comprehensive solution, there is a need for a holistic effort against misinformation, with users themselves being the \dquote{last line of defense}~\cite{Terren2023}.

A call for critical thinking has been put forth to improve information evaluation and opinion formation in increasingly chaotic online spaces~\cite{Plencner2014}. There has also been a call towards designing technologies that support critical thinking to enhance resistance towards misinformation~\cite{Boonprakong2023}. Responding to these, we present Iffy-Or-Not (\sys{}), a browser extension for reading online text that seeks to confer resistance towards misinformation.
\sys{} is informed by argumentation theory that examines how claims and conclusions can be supported or undermined during discourse through reasoning~\cite{Lewinski2016}. This includes identifying flaws in arguments~\cite{Aristotle} and attempts to persuade~\cite{Rapp2023}, and posing provocative questions to expose weaknesses in arguments~\cite{seeskin1987dialogue}.

\sys{} integrates techniques in argumentation theory and harnesses the extensive world knowledge and natural language reasoning abilities of LLMs to (A) identify common fallacies in the comments, (B) generate search queries to further probe the comments with, and (C) generate questions to elicit deeper consideration of the comments, thereby gearing users towards thinking more critically about the content they consume in real-time. A technical evaluation of the fallacy detection of the LLM we used showed that it had an average of 84\% accuracy. We included this value in the description of \sys{} to be transparent to users about its performance.

Through a user study (\N{18}) comprising a within-subjects experiment and interview where \sys{} was used in an online forum, we found that \sys{} leads users to be more attentive and cynical about the comments they encounter. For some users, it also enhances the search process by providing greater convenience in initiating search and easing the query generation process with suggested queries that align with what they have on their minds. However, we also note mixed reactions towards the user experience of the extension with participants finding that there were misses and inaccuracies by \sys{}. There were also misalignments with \sys{} on their information goals for the platform and their thinking process. Undesirable backfiring effects like information overload, the implied truth effect and overreliance on \sys{} were also raised as potential concerns for it. Nevertheless, they appreciated that \sys{} enhanced their critical evaluation of online content.

Our study makes the following contributions:

\begin{itemize}
    \item A set of design goals for supporting critical information evaluation of online texts, informed by argumentation theory and a formative study.
    \item A technical evaluation on the use of an LLM to classify fallacies commonly found in online discussions.
    \item A browser extension with features to support critical information evaluation, empirically evaluated through a user study.
\end{itemize}

\section{Related Work}

We situate our work in the literature on deliberation and misinformation, opinion formation, AI-assisted information evaluation and argumentation theory. We expand on these areas and describe design goals (DG) for critical information evaluation gleaned from the literature.

\subsection{Deliberation and Misinformation}

Social platforms have emerged as spaces for online deliberation~\cite{Wright2007, Halpern2013}. With widespread reach and accessibility, these information and communication technologies serve as digital public spheres for individuals from diverse backgrounds and geographical regions to consume information on and discuss societal issues~\cite{Semaan2014}.
With online commenting enabling greater dynamism, depth and diversity in online discussions, new knowledge and perspectives towards societal issues can be gleaned, driving the formation of better-informed opinions and discussions~\cite{Halpern2013}.

While these social platforms bring advantages to public deliberation, various factors present key challenges to them. The opinionated nature of social platforms manifests in the proliferation of unsubstantiated comments.
Crucially, the presence of misinformation undermines the integrity of the discussions by potentially causing misinformed opinions among the public~\cite{McKay2020}. Cases of disinformation campaigns being successful in influencing public opinion for political and financial purposes have been well documented in areas like national elections~\cite{Sharma2022} and public health~\cite{Balakrishnan2022}. The rise of powerful large language models that have the capacity to generate massive and diverse volumes of seemingly genuine and even persuasive content have escalated the potential threat of misinformation~\cite{DeAngelis2023}.

Various ways have been proposed to aid users in making better judgments on the veracity of the content they consume. This includes accuracy prompts that predispose them towards being more mindful of the veracity of the content~\cite{Pennycook2022}, inoculation interventions that expose users to pieces of misinformation or manipulation techniques to bolster resistance to misinformation~\cite{Traberg2022},
and labels that warn about the credibility of the content~\cite{Morrow2022}. While these have been shown to work, the measures are often employed before or after, rather than during the period of engaging with online content~\cite{Martel2023}. The informal and leisurely context in which people use social platforms may dampen their levels of analytical~\cite{Pennycook2019lazy} and effortful thinking~\cite{Zhong2011}, thereby diminishing the effects of these interventions. Furthermore, the effectiveness of such interventions has been found to fade over time unless constantly refreshed~\cite{Maertens2021}.
Technologies that support critical thinking have been called for to enhance the resistance towards misinformation~\cite{Boonprakong2023}. 

\subsection{Opinion Formation}

By reading about the views of others and the experiences they share, one can gain knowledge beyond what one can personally experience, and this lends towards informing one's opinion on an issue. Knowledge from the community, termed as the wisdom of the crowd~\cite{Surowiecki2004}, suggests that the collective intelligence pulled from a large pool of diverse and independent users can lead to more accurate portrayals of sentiments or outcomes than a small group of experts as aggregated opinions balance out individual biases. This is especially the case for controversial matters where facts are hard to establish or easily undermined, and where there are strong groups of ideologies with conflicting and competing interests.

While social platforms can provide a wealth of knowledge, users tend to only be presented with facets of content influenced by what they selectively consume~\cite{Cinelli2020} and are recommended with by algorithms running in the background~\cite{Bakshy2015}. Algorithms on social media often prioritize content that generates high engagement which can result in the amplification of sensational, polarizing, and fringe content~\cite{Vosoughi2018, Srba2023}. With the process of opinion formation being vulnerable to biases such as the availability heuristic where judgments are formed based on information that are readily available regardless of accuracy~\cite{Tversky1974}, this may result in the misplaced construction of one's worldview.

With social platforms having the capacity to impact opinion formation underscores the need to enhance the integrity of the content consumed online. A way to achieve this is by having users be more critical and discerning towards the content~\cite{Sadler2005, Machete2020}. Critical thinking encapsulates skills in \dquote{analyzing arguments, making inferences using inductive or deductive reasoning, judging or evaluating, and making decisions or solving problems}~\cite{Lai2011}. It includes being able to sift fact from fiction, gathering supporting and disputing evidence, and having a critical eye and healthy skepticism when evaluating information to form well founded opinions~\cite{Ku2019}. To improve the quality of online opinion formation, we thus came to \DG{1}{Encourage diverse and critical evaluation of content; foster an environment that stimulates scrutiny and engagement}.

\subsection{AI-Assisted Information Evaluation}

Evaluating information is an undertaking challenged by the overwhelming amount of content vying for one's attention. The endless slew of online content can lead to information overload, contributing to social fatigue~\cite{Zhang2022} and impeded information processing in terms of what information is prioritized and how quickly it is processed~\cite{GomezRodriguez2014}. 
This is further exacerbated by the pervasiveness of misinformation on the web~\cite{Zhou2020}.
This has led to a call for improving information evaluation and opinion formation in online spaces~\cite{Plencner2014}. 

As ways of mitigation, there has been growing work on the use of automated tools powered by artificial intelligence (AI) for the purposes of fact-checking. 
This spans from lightweight labeling interventions to end-to-end systems.
Generally, these works point towards the effectiveness of automated fact-checking. Labeling interventions have been observed to dampen undesirable behaviors such as reducing partisan bias~\cite{Moon2023}, and reducing the perceived accuracy~\cite{Sumpter2021, Jia2022, Lu2022} and sharing intention~\cite{Epstein2022} of false content. Fact-checking systems have also been shown to effectively assist users in verifying content~\cite{Lin2022, Karagiannis2020, Melo2019, Shin2022b}, even serving educational purposes~\cite{Karduni2019}. A caveat, however, is the implication of such a reliance on AI-assistance--that failures by the AI may backfire. For instance, a study found that users tended to adopt the AI's credibility indicators on posts, even when their veracity perceptions were actually correct and the AI's were erroneous~\cite{Lu2022}. A user study on a system that supported the assessment of the factuality of claims found that participants tended to trust the system even when the predictions were wrong, and the authors cautioned about overtrust in AI diminishing human accuracy~\cite{Nguyen2018}.
To harness the scalability and efficiency boost of automated technologies, yet also keep overreliance in check, we put forward \DG{2}{Avoid making veracity judgments with AI; assist users to reach a conclusion}.

There are existing automated tools that align with this design goal. Some works apply AI techniques to surface problematic aspects in online texts such as propaganda~\cite{Zavolokina2024}, unreliable and biased content~\cite{Horne2019}, and disputed content~\cite{Ennals2010}. These tools then provide additional information, such as explanations and related articles, to support the development of balanced perspectives. However, a gap remains between identifying potentially problematic content and guiding users through the reasoning process to reach an independent conclusion. LLMs have been employed to bridge this gap. LLMs can generate contextually relevant questions and provide answers to them by summarizing web documents, helping users find complementary information~\cite{Zhang2023, Quelle2024}. Chatbots have also been developed that use Socratic questioning to elicit deeper thinking about the content~\cite{Favero2024, Duelen2024}.
While numerous tools exist to detect problematic content and support user reasoning, they often address these aspects separately. Building upon these works, \sys{} unifies the information evaluation process, employing AI to enhance users' critical thinking when they engage with content they find intriguing or questionable. It encourages users to be more deeply involved in assessing content, empowering them to make their own judgments rather than deferring entirely to the AI.
\sys{} achieves this by leveraging fallacies, which provide the unique advantage of exposing logical flaws and baseless content, thereby steering users toward more thoughtful reasoning and a deeper search for reliable evidence.

\subsection{Argumentation Theory}

In deliberative and polarizing contexts, people often share their opinions to argue for their positions and persuade others to align with their views. Seeking to understand how this affects opinion formation led us to argumentation theory that has been a long-standing area of study in the field of philosophy. Argumentation theory looks into how people reason to defend their positions and challenge others' positions, especially in situations of disagreement or doubt~\cite{Lewinski2016}.
From the time of Socrates, the philosopher put forth a method of questioning where provocative questions are continually posed towards an interlocutor until a contradiction in their responses is exposed~\cite{seeskin1987dialogue}. This method has been regarded as the basis of critical thinking due to the \dquote{systematic, disciplined, and deep} mode of pursuing thought~\cite{Paul2016}. Later, Aristotle proposed three means of persuasion: ethos, pathos and logos, which refer to persuasion through establishing credibility, inciting emotions and reasoning with logic, respectively~\cite{Rapp2023}. These subsequently tied in with his work on fallacies~\cite{Aristotle}.

Fallacies are flaws in reasoning that can be categorized as formal, where an argument is invalid due to its logic, and informal, where an argument is invalid due to its content~\cite{Hansen2024}. Typically, insufficient or incorrect evidence is provided in claims where fallacies are present, making them \dquote{iffy}. Fallacies are common in online discussions given the opinionated nature of these social platforms~\cite{sahai2021breaking, Hidayat2020}. As fallacies are deceptive by nature and can often go unnoticed, researchers have considered their association with misinformation~\cite{Beisecker2024}. Musi and Reed identified ten fallacies that flagged for misinformation in news articles~\cite{Musi2022}. More topically focused, Cook surfaced several fallacies in climate change misinformation~\cite{Cook2022}, while Bonial et al. annotated five fallacies present in Covid-19 misinformation~\cite{bonial2022search}.

To enhance information evaluation in social platforms where algorithmic curation and human cognitive biases undermine the perceptiveness towards problematic content, researchers have called for technologies integrated with the affordances of these platforms to boost critical thinking~\cite{Boonprakong2023}. We answer to this call by incorporating knowledge from argumentation theory to develop \sys{} and conduct an assessment of it. By adopting argumentation theory, we seek to enhance user's critical evaluation as they read texts and assist them in probing deeper.

\section{Formative Study}

The goal of the formative study was to understand how participants' opinions morphed as they read social media comments in a deliberative setting and how nudging them towards thinking about the content more critically influenced their opinions. In doing so, we sought to illuminate more design goals for critical information evaluation.
We recruited 12 participants who were university students, with an equal number identifying as women and men.
Participation was voluntary and no compensation was provided. IRB approval was obtained for the study.

\subsection{Procedure}

The formative study involved two reading tasks and an interview with a duration of 1.5 hours. A demographics survey was taken at the start.

Participants first saw the titles of three posts taken from r/changemyview in Reddit. This subreddit was chosen due to its position of being a social media space appropriated for online deliberation~\cite{Tan2016}. The three threads were on automation replacing human workers\footnote{\url{https://www.reddit.com/r/changemyview/comments/14ppcrf/cmv_automation_and_ai_replacing_human_workers_is/}}, ex-convicts being treated unfairly in recruitment\footnote{\url{https://www.reddit.com/r/changemyview/comments/13vhh81/cmv_exconvicts_are_treated_excessively_unfairly/}}, and calling for those with genetic disorders to abstain from reproducing\footnote{\url{https://www.reddit.com/r/changemyview/comments/14lmx8c/cmv_people_who_suffer_from_a_genetic_dissorder/}}. Participants selected the post that they had the strongest opinion about (either in agreeing or disagreeing) and indicated their stance from 1: strongly disagree to 7: strongly agree.

In each reading task, participants were told to think aloud as they read the selected thread.
There were 7 comments shown for each thread that were randomly ordered. The two tasks followed a pre-post experiment design where participants first read and rated the thread, then read a primer on various aspects of argumentation theory, and were asked to read and rate the thread again. The primer had 5 slides on evaluating the strength of arguments, and the modes and misuses of persuasion.

The interview sought to understand how participants perceived the comments before and after the primer, and how it affected their stance and evaluation of the comments.
Participants also reflected on their experiences with social media platforms.

\subsection{Results}

Five participants chose the automation thread (2 disagree; 3 agree), three chose ex-convicts (1 disagree; 2 agree) and four chose the thread on genetic disorders (3 disagree; 1 agree). The brackets indicate their stances after reading solely the title of the thread.
The ratings on the stances towards the thread and opinion changes from the comments are statistically analyzed with descriptive and inferential statistics reported. The interviews were thematically analyzed by surfacing broad patterns and unique instances to glean deeper insights~\cite{Braun2006}.

\subsubsection{Stances on the Thread}

Participants' stances were measured another three times: after reading the post by the user who started the thread, after reading all the comments (pre-reading), and after reading all the comments again after the primer (post-reading). A significant difference was observed in the stances between reading the user's post and the pre-reading of the comments (\Wilcox{-50}{=.0369}). This suggests that the comments impacted the opinion participants formed of the issue after gaining more knowledge from them. Between the pre-reading and post-reading, only three participants adjusted their stance which retained the same polarity and were just slightly stronger or weaker by one degree.

\subsubsection{Interview Insights}

The interviews provided illuminating insights into the participants' behaviors and experiences, and we drew on them to identify the design goals for critical information evaluation.

\paragraph{Opinion formation} When it comes to evaluating the comments, participants mentioned that they were seldom inclined to change their views, perhaps a factor being that we initially asked them to select the topic that they already felt the strongest about and was thus resistant to change~\cite{Zuwerink1996}. They said that they were more open to change when logical comments, concrete evidence and examples, or information that revealed a flaw in their current understanding were presented. Comments that reinforced their view typically aligned with what they believed, providing objective evidence and new perspectives. Comments that participants regarded to not influence them and tended to dismiss were emotionally charged, inflammatory, purely rhetorical, or generally short and uninformative as have been observed in prior work~\cite{Tormala2002}. Anecdotal evidence, in particular, brought about split views as some participants dismissed them whereas others valued them in offering a testimony that they could relate to. Overall, participants showed a preference for objective and informative content over opinionated and unhelpful content. These observations brought us to \DG{3}{Offer objective and informative content; encourage curiosity and streamline discovery}.

When asked about how their reading changed after the primer, many participants mentioned that they felt more critical and neutral. For the former, they looked for innuendos of persuasion in what and how the author was writing, and whether reliable evidence was provided. For the latter, they mentioned detaching themselves to reconsider more emotionally charged comments without their emotional appeal. While some participants favored being more attentive and critical in the post-reading, the issue of cognitive burden was frequently raised. Participants mentioned that it was a more challenging and tedious approach to reading the comments.
Actively interpreting and filtering information is an effortful process constrained by the limited bandwidth of human cognition~\cite{Miller1957}. By taking such load off humans through automated means, cognitive reserves can be dedicated to more impactful activities like verifying the information and acquiring knowledge~\cite{Maedche2019}. This led us to \DG{4}{Alleviate cognitive load and channel to purposeful tasks; signal for potentially problematic content}.

\paragraph{Reading Social Media Comments} Expanding into their real social media experiences, participants mentioned points similar to the above on the types of comments they picked out or glossed over when they were trying to have a better grasp of the discussion topic. They appreciated comments that provided background or offered fresh perspectives. They mentioned dismissing comments that were negative, toxic or reductive like those with name-calling or straw man arguments, mentioning that such comments stressed and offended them.

When asked about their encounters with misinformation, those who have experienced them come to realize the misinformation because they were already familiar with the topic and the facts about them, corroborating prior work~\cite{Pomerantz1996}. They also mentioned realizing the misinformation by seeing comments from others who have called them out. A participant also brought up how human judgment is prone to biases in terms of what is written and how the content is read and that it is hard to gauge these biases. When they are suspicious about the content and keen to find out more, some participants mentioned doing web searches or asking trusted persons, whereas some rode on the collective intelligence afforded by these platforms by going through the replies and quotes or waiting for time to pan out so that the community can offer a public opinion. These remarks indicate how a lack of knowledge and inherent heuristics and prejudices may cloud judgment, which underscores \DG{1}{Encourage diverse and critical evaluation of content; foster an environment that stimulates scrutiny and engagement}.

\section{Iffy-Or-Not (ION)}

From the literature review, we surfaced four design goals for critical information evaluation:

\begin{enumerate}
    \item Encourage diverse and critical evaluation of content; foster an environment that stimulates scrutiny and engagement
    \item Avoid making veracity judgments with AI; assist users to reach a conclusion
    \item Offer objective and informative content; encourage curiosity and streamline discovery
    \item Alleviate cognitive load and channel to purposeful tasks; signal for potentially problematic content
\end{enumerate}

Aligning with aspects of argumentation theory and prior work, we conceived three features for \sys{}: highlight, probe and chat (Figure~\ref{fig:system}). 
The highlight feature integrates DG4 and theories of persuasion and fallacies to identify and raise to attention fallacious content. The probe feature considers DG1 and DG3 by suggesting queries and summarizing content from the web to examine the content from several angles and sources. And the chat feature combines DG1 and the Socratic Method to provide questions that surface different perspectives to interrogate the content with. DG2 underlies these three features which guide users towards assessing and forming their own perceptions of the reliability and usefulness of the content they consume without imposing any veracity judgements--the nature of fallacies suggests failure to prove a point and not that it is false.

We operationalized LLMs to power these features. LLMs have strong natural language reasoning and generation capabilities that are suitable for knowledge-intensive tasks~\cite{Yang2024}. Trained on large amounts of data, they also have an extensive knowledge base that makes them suitable for our purposes of critical reading and evaluation that can encompass a broad range of topics.
To emphasize to users that the features are powered by LLMs, we named the extension they used by \textit{AI-Checker} in place of \textit{Iffy-Or-Not} to make it more intuitive for users to know that the extension was powered by \textit{AI}, which is a more ubiquitous term than LLM, and that it served to \textit{check} texts for fallacious content.

\subsection{Selection of the Fallacies}

To establish a set of fallacies for \sys{}, we referred to work that classified, identified or used fallacies in the context of online discussions~\cite{sahai2021breaking, Hidayat2020, nikolaidis2023experiments} and misinformation~\cite{Musi2022, Beisecker2024, Hruschka2023, Cook2018, bonial2022search, jin2022logical, Sourati2023, Lundy2023} and selected those that were the most commonly mentioned (\textit{against the person}, \textit{appeal to authority}, \textit{appeal to popularity}, \textit{appeal to emotion}, \textit{questionable cause}, \textit{hasty generalization}, and \textit{red herring}). Based on university educational resources~\cite{Kashyap2020, GGU2023}, the fallacies were further categorized into the persuasive strategies (\textit{ethos}, \textit{pathos}, and \textit{logos}) theorized in Aristotle's Rhetoric~\cite{SEP2022}. Table~\ref{tab:logfal} shows the categorization and definitions of the fallacies. From a technical evaluation following the same protocol described in Section~\ref{sec:techeval}, hasty generalization and red herring were removed from the set of fallacies due to the subpar performance of the LLM.

\begin{table*}[htb!]
  \caption{Common fallacies found in online discussions and misinformation categorized by persuasive strategies.}
  \label{tab:logfal}
  \Description{A table showing the five fallacies and their definitions categorized by their persuasive strategy. They are against the person and appeal to authority for ethos, appeal to popularity and appeal to emotion for pathos and questionable cause for logos.}
  \begin{tabular}{ccp{.55\textwidth}}
    \toprule
    \makecell{Persuasive\\Strategy} & \makecell{Fallacy\\(in Latin)} & Definition\\
    \midrule
    Ethos & \makecell[t]{Against the Person\\(Argumentum Ad Hominem)} & Attacking the person or some aspect of the person making the argument instead of addressing the argument directly.\\
     & \makecell[t]{Appeal to Authority\\(Argumentum Ad Verecundiam)} & Using an alleged authority who is not really an authority on the facts relevant to the argument as evidence.\\
    \midrule
    Pathos & \makecell[t]{Appeal to Popularity\\(Argumentum Ad Populum)} & Affirming that something is real or better because the majority in general or of a particular group thinks so.\\
    & \makecell[t]{Appeal to Emotion\\(Argumentum Ad Passiones)} & Manipulating the reader's emotions in order to win the argument in place of a valid reason.\\
    \midrule
    Logos & \makecell[t]{Questionable Cause\\(Non Causa Pro Causa)} & Concluding that one thing caused another simply because they are regularly associated.\\
    \bottomrule
  \end{tabular}
\end{table*}

\subsection{Features of the Extension}

\begin{figure*}[!htb]
    \centering
    \begin{subfigure}[b]{.3\textwidth}
        \centering
        \includegraphics[width=\textwidth]{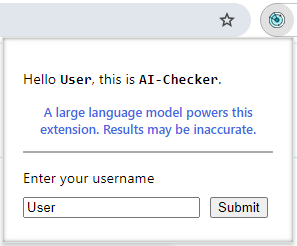}
        \caption{Access segment}
        \label{fig:extension1}
    \end{subfigure}
    \hspace{.03\textwidth}
    \begin{subfigure}[b]{.3\textwidth}
        \centering
        \includegraphics[width=\textwidth]{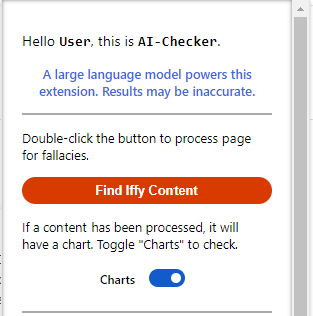}
        \caption{Function segment}
        \label{fig:extension2}
    \end{subfigure}
    \hspace{.03\textwidth}
    \begin{subfigure}[b]{.3\textwidth}
        \centering
        \includegraphics[width=\textwidth]{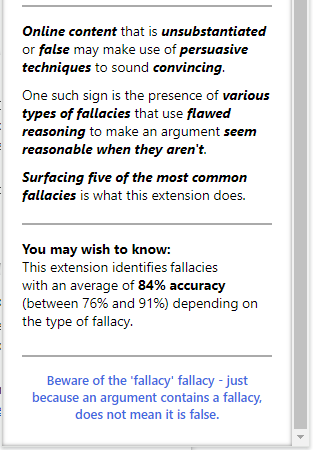}
        \caption{Description segment}
        \label{fig:extension3}
    \end{subfigure}
    \caption{The \sys{} extension. The access segment allows the user to enter their preferred username. The \dquote{Find Iffy Content} button in the function segment runs the fallacies detection. The description segment explains what the extension does.}
    \Description{Three figures showing the extension page of Iffy-Or-Not. The first has a username field. The second has a warning on LLM inaccuracies and a red button to activate the extension. The third has a description on what the extension does, a statement on it's accuracy, and a statement on the Fallacy fallacy.}
\end{figure*}

\sys{} is a Chrome extension that highlights fallacious content in a web page for users to probe deeper into and chat about. It activates by clicking the extension icon. This makes a popup appear that asks the user to enter a username of their choice (Figure~\ref{fig:extension1}). Thereafter, it shows the \dquote{Find Iffy Content} button (Figure~\ref{fig:extension2}) and descriptions of what \sys{} does (Figure~\ref{fig:extension3}). Clicking the \dquote{Find Iffy Content} button activates the features of \sys{}.
The three features of \sys{} are interconnected and explained in Figure~\ref{fig:system}.

The highlight feature informs the user about claims they may want to pay attention to as those are fallacious and likely unsubstantiated. There is a chart that summarizes all the fallacies detected and gives their definitions in an overlay on the corresponding part of the chart that is hovered upon (\hchart{};~A1i). When the user hovers over the highlights (\hai{};~A1ii), tooltips appear that describe what fallacy was detected and the explanation for the fallacy given by \sys{} (\hsays{};~A1iii). Since there are five fallacies, each one was assigned a color from the rainbow except red. Content marked by users were indicated by a light red underline with the reasons also viewable as tooltips (\huser{};~A2).

The probe feature offers three search queries for each highlighted content generated by \sys{} (\psug{};~B1i) that opens a box showing a summary of three findings drawn from the web for the selected query (\pweb{};~B1ii). The user can also enter their own search terms to generate another set of queries (\pwrite{};~B2). Selecting a query starts a Google search in a new tab that they can browse through.

The chat feature provides suggestions by \sys{} on how to critically evaluate the highlighted content (\cfood{};~C1). The suggestions can be refreshed to show new questions. There is also a field for users to enter their messages to chat with other users and vote on their messages (\cdis{};~C2). Having a chat for each highlight provides a space for situated discussions. The colors of the headers correspond to the detected fallacy.

\subsection{Technology Stack}

\begin{figure*}[!htb]
    \centering
    \begin{subfigure}[b]{.3\textwidth}
        \centering
        \includegraphics[width=\textwidth]{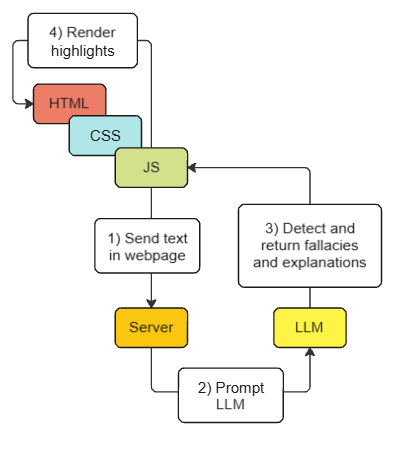}
        \caption{Highlight}
        \label{fig:stack1}
    \end{subfigure}
    \hspace{.03\textwidth}
    \begin{subfigure}[b]{.3\textwidth}
        \centering
        \includegraphics[width=\textwidth]{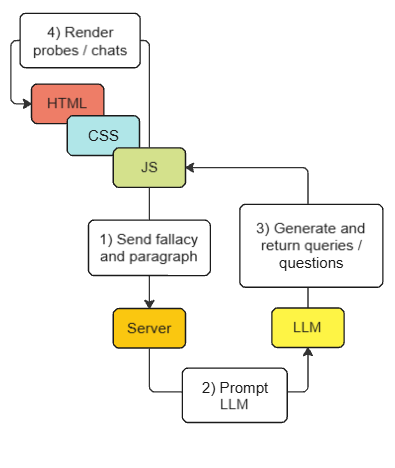}
        \caption{Probe queries / Chat questions}
        \label{fig:stack2}
    \end{subfigure}
    \hspace{.03\textwidth}
    \begin{subfigure}[b]{.3\textwidth}
        \centering
        \includegraphics[width=\textwidth]{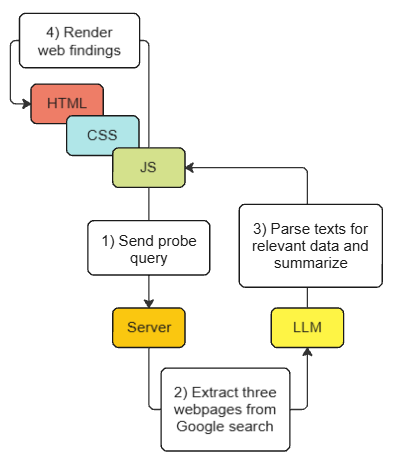}
        \caption{Probe web findings}
        \label{fig:stack3}
    \end{subfigure}
    \caption{Technology stack for each feature of \sys{}.}
    \Description{Three figures showing the technology stacks for various features of Iffy-Or-Not. All stacks are similar in that content is passed to the server who prompts the LLM that returns the response back so that HTML, CSS and JavaScript can be used to populate the webpage with the desired content.}
    \label{fig:stack}
\end{figure*}

\sys{} was implemented as a Chrome browser extension\footnote{\url{https://developer.chrome.com/docs/extensions}}. The client side was developed using basic HTML, CSS and JavaScript. A Flask\footnote{\url{https://flask.palletsprojects.com/en/3.0.x/}} server, developed with Python, made API calls to a locally hosted Meta-Llama-3-8B-Instruct\footnote{\url{https://huggingface.co/meta-llama/Meta-Llama-3-8B-Instruct}} LLM. Cloud Firestore\footnote{\url{https://firebase.google.com/docs/firestore}} was used as the database to capture interactions on the extension. Figure~\ref{fig:stack} shows how these interacted for each feature. Generally, the client passed relevant content from the web to the server that would provide the appropriate prompt (see Appendix~\ref{sec:prompts}) to the LLM. The response by the LLM is then passed back to the client to render the new content.

As \sys{} hinges on the performance of the detection of fallacies, we describe a technical evaluation of the LLM used for our study following prior work~\cite{Lim2024}. We did not evaluate the other prompts as most of them were for creative purposes (producing queries and questions) rather than decision-making (classifying fallacies). 

\subsection{Technical Evaluation of Fallacies Detection}
\label{sec:techeval}

To evaluate the LLM, we used the LOGIC dataset by Jin et al.~\cite{jin2022logical} which comprises 2,449 examples of fallacies across 13 categories. We wrote multiple queries to filter the data, excluding those that (a) were duplicates, (b) fell outside the scope of our fallacies, (c) only defined or described fallacies, (d) included Latin phrases, and (e) contained quiz-related phrases. The last was due to the LOGIC dataset being sourced from student quiz websites. We also manually reviewed the data to remove cases overlooked by the queries. This yielded a reduced dataset of 546 instances with the breakdown as follows: 160 against the person, 77 appeal to authority, 136 appeal to popularity, 44 appeal to emotion and 129 questionable cause.
A dataset comprising statements of scientific facts from a science website\footnote{\url{https://www.jameswebbdiscovery.com/universe/100-interesting-facts-about-life-unraveling-the-wonders-of-existence}} that was unlikely to contain fallacies was further included, with the 99 instances labeled as \nth{}.
Altogether, the dataset had 645 instances.

The \emph{Meta-Llama-3-8B-Instruct} model\footnote{\url{https://huggingface.co/meta-llama/Meta-Llama-3-8B-Instruct}} was used to identify fallacies. It was set to a temperature of \texttt{0}, with a maximum number of \texttt{512} new tokens, and the role: \texttt{You are a critical thinker.} Few-shot prompting~\cite{Brown2020} was used for the fallacy classification task, with three examples given for each of the five fallacies. These 15 instances were taken from the dataset and therefore excluded in the evaluation of the LLM's performance (\N{630}). The prompt is provided in Appendix~\ref{app:detectprompt}.

The multi-class classification performance metrics~\cite{grandini2020metrics} of the LLM in identifying fallacies is shown in Table~\ref{tab:fullmetrics}. If a predicted label did not belong to the classification set, we modified it to \nth{} (\n{2}). When prompting, we also allowed the LLM to answer \nth{} if no fallacy was identified in a text (\n{109}). As such, we also present the performance metrics on the subset (\N{519}) of the data, where 111 instances predicted as \nth{} were removed. The normalized classification matrices for both datasets are shown in Appendix~\ref{app:cmatrix}.

\begin{table}[!htb]
  \caption{Multi-class classification performance of the LLM in identifying fallacies for the full data (\N{630}) and the subset data (\N{519}).}
  \label{tab:fullmetrics}
  \Description{A table showing the accuracy of the LLM where the accuracy for the full dataset is 85 percent and accuracy for the subset is 84 percent.}
  \begin{tabular}{cccccc}
    \toprule
    Data & Accuracy & Averaging & Precision & Recall & F1 Score\\
    \midrule
    Full & 0.85 & Macro & 0.82 & 0.83 & 0.82\\
    Subset & 0.84 & Weighted & 0.87 & 0.85 & 0.85\\
    \bottomrule
  \end{tabular}
\end{table}

The LLM achieved an average accuracy of 0.85 in classifying the fallacies (Figure~\ref{fig:ncm_full}). Most of the inaccuracies occurred in failing to identify the appeal to popularity fallacy, with a 27.1\% misclassification rate (Table~\ref{tab:breakdown}). There was also more to be desired with the LLM in identifying the appeal to emotion (24.4\% misclassification) and appeal to authority (23.0\% misclassification) fallacies.
The difficulty that LLMs have in extracting fine-grained structured sentiment and opinion information~\cite{zhang2023sentiment} may account for the poor performance in identifying these fallacies, which require more nuanced interpretations of texts.
The LLM performed best for the against the person fallacy with a 8.92\% misclassification rate. Interestingly, cases of hallucination were observed with the LLM, where it offered fallacies outside our considered set, such as \dquote{appeal to tradition} and \dquote{appeal to celebrity}.

\begin{table*}[!htb]
  \caption{Breakdown of the instances in the full data (\N{630}) that were classified as \nth{} or misclassified as another label including none. Percentages are calculated against the `All' row.}
  \label{tab:breakdown}
  \Description{A table showing the instances of the detected fallacies. In total, there are 531 instances, of which 2.54 percent was classified as Nothing and 14.8 percent was misclassified.}
  \begin{tabular}{cccccc|c}
    \toprule
    & \makecell{Against the\\Person} & \makecell{Appeal to\\Authority} & \makecell{Appeal to\\Popularity} & \makecell{Appeal to\\Emotion} & 
    \makecell{Questionable\\Cause} & Total\\
    \midrule
    All & 157 & 74 & 133 & 41 & 126 & 531\\
    \midrule
    Classified as \nth{} & 0 & 2 & 5 & 6 & 3 & 16\\
    Percentage & 0 & 2.70 & 3.76 & 14.6 & 2.38 & 2.54\\
    \midrule
    Misclassified & 14 & 17 & 36 & 10 & 16 & 93\\
    Percentage & 8.92 & 23.0 & 27.1 & 24.4 & 12.7 & 14.8\\
    \bottomrule
  \end{tabular}
\end{table*}

When considering the subset data in which \nth{} instances were removed, the LLM had a similar accuracy of 0.84, with the classification accuracy of the fallacies ranging from 0.76 to 0.91 (Figure~\ref{fig:ncm_subset}). Given that our system does not require surfacing \textit{all} fallacies, but should classify them \textit{accurately}, we consider the LLM capable for our task definition. Going forward, we made the 84\% accuracy of the LLM in identifying fallacies in \sys{} transparent to the users.

\section{Method}

To assess \sys{}, we conducted a user study that includes a pre-survey, experiment, post-survey and interview.
The duration of each session was 2 hours and participants were paid \textasciitilde{}USD 15 for their time. IRB approval was obtained for the study.

\subsection{Pre-Survey}

At the start, participants filled a pre-survey on their demographics, and their familiarity with social media, LLM tools, and fallacies. We asked for their ease of discussing the five topics used for the study as knowledge is a factor of critical thinking~\cite{Lai2011}. Informational awareness~\cite{Tandoc2021} was measured as misinformation susceptibility is affected by media literacy~\cite{AdjinTettey2022, Lu2024}. We also measured the need for cognition~\cite{LinsdeHolandaCoelho2020} as it influences both critical thinking~\cite{West2008} and the susceptibility to misinformation~\cite{Xiao2021, Wu2023}.

\subsection{Experiment}

Before the experiment, we showed participants a 1-minute video explaining the think aloud protocol~\cite{Lewis1982}. Thereafter, we gave them a walkthrough of each feature of \sys{} using a Reddit thread on climate change\footnote{\url{https://www.reddit.com/r/changemyview/comments/paqlx8/cmv_everyday_people_will_have_to_make/}}. They could read any user highlights or chat messages left by previous participants. We briefly collected their thoughts towards each feature as we went through them.

Four threads were used for the experiment that were a mix of several topics. The threads were on population decline\footnote{\url{https://www.reddit.com/r/changemyview/comments/1eun5sv/cmv_population_decline_is_self_medicating_problem/}}, weight loss\footnote{\url{https://www.reddit.com/r/changemyview/comments/7kfu0e/cmv_popular_culture_should_stop_spreading_the/}}, prohibiting web access for children\footnote{\url{https://www.reddit.com/r/changemyview/comments/vdscsa/cmv_kids_under_13_should_be_legally_prohibited_to/}}, and AI taking over jobs\footnote{\url{https://www.reddit.com/r/changemyview/comments/1bj1ddz/cmv_ai_is_going_to_take_all_jobs_soon_which_will/?rdt=58371}}. The posts were between 300 to 350 words. We ensured that each post had four AI highlights for consistency.

Prior work assessed related systems~\cite{Xia2022, Kim2024, Zhang2023} using experiments that had simple yet relevant tasks that are aligned with potential use cases in real-world scenarios to encourage participants to use the full functionalities offered by the system in a controlled, measurable, and comparable setting. Adopting their approach, our experiment had a within-subjects design where participants would read two of the four Reddit threads, with and without \sys{}, randomized using Latin square design. The task was inspired by X's community notes in which users provide more context to posts that might be misleading\footnote{\url{https://communitynotes.x.com/guide/en/about/introduction}}. The task description was as follows, where the text in < > was removed for the non-\sys{} task:

\begin{displayquote}
You are a user who is critical about the content posted on Reddit. You will look through the main post and pick two highlights that you think are the most dubious. You will then write a message in each of their chats that aims to better inform future readers who come across this post. Your message should be well-substantiated to support or refute the highlighted content. You <will use the AI-Checker (highlight, probe, chat features) and> can use the web browser in any way you want to as well. Think aloud as you do so. You will have 15 minutes (with reminders midway and towards the end).
\end{displayquote}

Participants were asked to think aloud at all times--most gradually fell silent and we did not interrupt them if they did so. They could also use the web browser freely. The attached researcher would observe them throughout, taking notes on questions to follow up on during the interview. After each task, participants completed the NASA Task Load Index (NASA-TLX)~\cite{Hart1988} as a measure of cognitive load and rated their familiarity with the topic and level of satisfaction with what they wrote.

\subsection{Post-Survey}

The post-survey incorporated items from the \dquote{Information Quality} and \dquote{Service Quality} measures of the DeLone and McLean Information Systems Success Model~\cite{Delone2003} based on prior work~\cite{Tona2012, Alzahrani2019} with slight modifications to fit the context of this study. On the \dquote{System Quality} measure of the model, the System Usability Scale (SUS)~\cite{Brooke1996} was administered instead. We also asked participants to rate how likable and useful they found each of the three \sys{} features and to rank them in order of preference. For consistency with the SUS, all items were measured on a 5-point Likert scale.

\subsection{Interview}

A semi-structured interview was conducted at the end. The first half covered observations recorded by the researcher during the experiment. Participants were reminded about what they did and asked to clarify the reasons for their actions. The second half asked about their views and experiences with \sys{} during the walkthrough and the tasks. We sought to understand the differences they felt between the two tasks and whether having \sys{} aided them. We also considered if their prior knowledge on the topic had an impact. We then delved into their views toward each feature in \sys{}, its aesthetics and usability, and whether they would consider having such an extension in reality.

\subsection{Participants}

We recruited 18 participants by posting a call for participation in a Telegram channel for research studies in the region and via an email distribution list of a university. There was an equal number of women and men participants with a majority identifying as ethnically Chinese (\n{11}). They were aged between 18 to 24 (\n{4}), 25 to 34 (\n{9}) and 35 to 44 (\n{5}) years old. Participants were generally highly educated, having attained a polytechnic diploma (\n{2}), Bachelor's degree (\n{10}) and Master's degree (\n{6}). All participants were proficient in English.

The following items were measured on a 5-point Likert scale.
Participants were somewhat familiar with social media (\MSD{3.44}{0.69})\footnote{Familiarity with social media was averaged across familiarity towards six common social media platforms on a 5-point Likert scale.}.
They used LLM technologies like ChatGPT frequently (\MIQR{4}{0.75}), but less so for fact-checking purposes (\MIQR{3}{2}).
They reported a somewhat high level of informational awareness (\MSD{3.72}{0.69})\footnote{Informational awareness was averaged across familiarity towards 3 items on a 5-point Likert scale.}.
They had lower need for cognition (\MSD{3.28}{0.73}) than the average of British or American counterparts~\cite{PsyToolkit2023}.
Participants expressed some familiarity with fallacies (\MIQR{3}{2}).

On a 7-point Likert scale, they reported their ease of having a discussion about the five topics used for the study: climate change (\MIQR{5}{1}), AI taking over jobs (\MIQR{5}{1}), prohibiting web access for children (\MIQR{5}{2}), population decline (\MIQR{5}{1.75}), and weight loss (\MIQR{6}{1}).

\section{Results}

For the quantitative results, all items were measured on a 5-point Likert scale unless otherwise specified. Paired samples t-test~\cite{Student1908} was used for normally distributed data and Wilcoxon signed-rank test~\cite{Wilcoxon1945} for non-normal data.

\subsection{Tasks}
\label{sec:tasks}

\paragraph{Cognitive Load}

From the NASA-TLX, participants experienced lower cognitive load when using \sys{} (\MSD{60.2}{19.2}) for the tasks compared to without it (\MSD{64.7}{19.8}). The difference was not significant (\ttest{1.21}{=.244}).

\paragraph{Interactions}

During the \sys{} task, participants interacted the most with the probe feature (\n{100}), followed by the chat feature (\n{92}), and highlight feature (\n{31}). They also frequently used the browser (\n{83}).
Table~\ref{tab:interactions} shows the various interactions participants had with each feature in \sys{} and with the browser.

\begin{table}[!htb]
  \caption[Means (M) and standard deviations (SD) of the number of interactions per user for each of the features of \sys{} and the browser, split by the task conditions. Browser interactions were mainly on Google Search. As Google Search's AI Overview feature was released during the study period which earlier participants (P1-12) did not have access to, the interactions were combined with Featured Snippets that appear at the top of the results page instead. People Also Ask are groups of featured snippets that appear midway through the results page.]{Means (M) and standard deviations (SD) of the number of interactions per user for each of the features of \sys{} and the browser, split by the task conditions. Browser interactions were mainly on Google Search\footnotemark{}. As Google Search's AI Overview\footnotemark{} feature was released during the study period which earlier participants (P1-12) did not have access to, the interactions were combined with Featured Snippets\footnotemark{} that appear at the top of the results page instead. People Also Ask\footnotemark{} are groups of featured snippets that appear midway through the results page.}
  \label{tab:interactions}
  \Description{Four tables showing the means and standard deviations of the number of interactions per user for each of the features of Iffy-Or-Not and the browser, split by the task conditions. Browser interactions were mainly on Google Search. As Google Search's AI Overview feature was released during the study period which earlier participants (P1-12) did not have access to, the interactions were combined with Featured Snippets that appear at the top of the results page instead. People Also Ask are groups of featured snippets that appear midway through the results page.}
  \begin{tabular}{ccccc}
    \toprule
    & \multicolumn{2}{c}{\sys{}} & \multicolumn{2}{c}{\none{}} \\
    \cline{2-5}
    Highlight & M & SD & M & SD \\
    \midrule
    Summary Chart & 0.17 & 0.71 & - & - \\
    AI Highlight & 1.39 & 1.46 & - & - \\
    User Highlight & 0.17 & 0.51 & - & - \\
    \bottomrule
    \\
  \end{tabular}
  
  \begin{tabular}{ccccc}
    \toprule
    & \multicolumn{2}{c}{\sys{}} & \multicolumn{2}{c}{\none{}} \\
    \cline{2-5}
    Chat & M & SD & M & SD \\
    \midrule
    Food-For-Thought & 2.94 & 1.63 & - & - \\
    Discussion Space & 2.17 & 1.15 & 2.06 & 1.39 \\
    \bottomrule
    \\
  \end{tabular}

  \begin{tabular}{ccccc}
    \toprule
    & \multicolumn{2}{c}{\sys{}} & \multicolumn{2}{c}{\none{}} \\
    \cline{2-5}
    Probe & M & SD & M & SD \\
    \midrule
    Suggested Queries & 1.50 & 0.92 & - & - \\
    Web Findings (WF) & 1.56 & 1.65 & - & - \\
    WF: Open Reference & 0.50 & 0.79 & - & - \\
    Write Own Query (WOQ) & 1.00 & 1.53 & - & - \\
    WOQ: Open Query & 1.06 & 1.70 & - & - \\
    \bottomrule
    \\
  \end{tabular}
  
  \begin{tabular}{ccccc}
    \toprule
    & \multicolumn{2}{c}{\sys{}} & \multicolumn{2}{c}{\none{}} \\
    \cline{2-5}
    Browser & M & SD & M & SD \\
    \midrule
    Search Google & 1.50 & 1.20 & 2.89 & 2.17 \\
    View Featured Snippets & 1.17 & 1.50 & 0.89 & 1.02 \\
    View People Also Ask & 0.94 & 2.48 & 0.89 & 1.64 \\
    Open Link & 1.11 & 1.13 & 2.61 & 2.75 \\
    \bottomrule
  \end{tabular}
\end{table}

\footnotetext{\url{https://www.google.com/}}
\footnotetext{\url{https://blog.google/products/search/generative-ai-google-search-may-2024/}}
\footnotetext{\url{https://blog.google/products/search/reintroduction-googles-featured-snippets/}}
\footnotetext{\url{https://developers.google.com/search/docs/appearance/visual-elements-gallery\#related-questions-group}}

As the goal of \sys{} was to assist in the critical reading and evaluation of content, we aggregated the instances in which participants initiated searches (Suggested Queries, Write Own Query and Search Google, see Table~\ref{tab:interactions}) and viewed the search results (Web Findings, WF: Open Reference, WOQ: Open Query, View Featured Snippets, View People Also Ask, and Open Link) within \sys{} and in the browser. We found significantly more (\Wilcox{11}{=.007}) participants initiating searches in the \sys{} task (\MSD{4.06}{2.41}) than in the \none{} task (\MSD{2.89}{2.17}). Significantly more (\ttest{-2.15}{=.047}) participants viewed search results in the \sys{} task (\MSD{6.32}{4.56}) than the \none{} task (\MSD{4.39}{3.16}) as well.

\paragraph{Usage Patterns}

To understand how participants used \sys{}, we describe our observations during the task.

For the highlight feature, participants paid attention to the \hai{}s, which was expected of the task. Highlighted content was thought to be potentially problematic in general, rather than viewed as specific fallacies. Surprisingly, few participants looked at the \hsays{}s. Interviews revealed that participants had already become familiar with the fallacies during the walkthrough that took place earlier and did not find the need to refer any further.

For the probe feature, participants tended to prioritize \psug{}. Most participants opened \pweb{} in a targeted manner, using it for information-seeking. A few participants opened many \pweb{} simultaneously, suggesting that they used it for knowledge-building instead.
When using \pwrite{}, participants who entered long phrases often remarked that the outputs were irrelevant and proceeded to use Google Search, while those who entered short phrases were pleased by the results. This former is likely because phrases such as \dquote{is there evidence that people who exercise more will increase their regular intake?} are typically complete questions. Yet, \pwrite{} is designed to offer diverse suggestions that can at times deviate from the input.

For the chat feature, there was varied use of \cfood{}. Those who viewed more questions either struggled to decide what to write or sought additional aspects to include in their message. For instance, a participant saw the \cfood{}: \dquote{Is the author's tone \ldots{} respectful?} and stated, \dquote{This is definitely a stereotype and generalization \ldots{} the author is not giving respect by making such assumptions.} before incorporating this perspective in their message. Participants who engaged less with \cfood{} either prioritized looking up information or writing their messages.

\paragraph{Written Messages}

Participants wrote about two messages for each task (see Discussion Space in Table~\ref{tab:interactions}), and were similarly satisfied with the messages in either conditions (\sys{}: \MIQR{4}{1}; \none{}: \MIQR{4}{1}). Upon reviewing the messages, we observed that participants took the same approach in crafting them. For instance, if the messages for the \sys{} task had the layout of indicating their personal opinion first followed by inserting links from the web, the messages for the \none{} task were similar. The verbosity and level of details in the messages such as citing statistics also followed a similar pattern.

This is an example of a message by a participant (P6) for the \sys{} task:
\begin{displayquote}
\textit{Hysterical statement. Assumes that the transition to AI will truly lead to massive unemployment. Not quite what the experts think. \url{https://think.ing.com/articles/ai-will-fundamentally-transform-job-market-but-risk-of-mass-unemployment-is-low/}}
\end{displayquote}

And this is an example of a message by the same participant for the \none{} task:
\begin{displayquote}
\textit{Oversimplistic comment. Curbing your caloric intake is indeed important to weight loss, but it is certainly not the only factor as obesity is a complex syndrome. \url{https://www.nhlbi.nih.gov/health/overweight-and-obesity/causes}}
\end{displayquote}

\subsection{Post-Survey}
\label{sec:postsurvey}

\paragraph{Information Quality}

Participants felt that \sys{} was able to provide the type of information they needed (\MIQR{4}{0}), and to a sufficient amount (\MIQR{4}{1}). It also presented information in a clear and understandable way (\MIQR{4}{0}), and they were satisfied with its overall accuracy (\MIQR{4}{0}).

\paragraph{Service Quality}

While they did not find \sys{} to be very visually appealing (\MIQR{3}{1}), they were satisfied with their interactions with it (\MIQR{4}{0.75}).

\paragraph{System Quality}

The SUS scores (\MSD{74.3}{9.11}) indicated that the usability of \sys{} was slightly above average, with the average being 68.1 for web-based interfaces~\cite{Bangor2008}.

\paragraph{Perceptions of Each Feature}

Participants rated liking each feature: highlight (\MIQR{4}{0}), probe (\MIQR{4}{1}), chat (\MIQR{4}{0.75}). They also found them useful: highlight (\MIQR{4}{0.75}), probe (\MIQR{4}{1}), chat (\MIQR{4}{1}).
When asked to rank the features in order of preference, probe came in at the top (\MSD{2.11}{0.83}), followed by chat (\MSD{2.06}{0.80}), then highlight (\MSD{1.83}{0.86}). The scores ranged from 1 to 3, where 3 refers to the feature being the most preferred.

\subsection{Assessing the Helpfulness of \sys{}}
\label{sec:helpfulness}

From the literature, critical thinking is influenced by knowledge (\ed{})~\cite{Lai2011}, misinformation susceptibility is influenced by media literacy (\ia{})~\cite{AdjinTettey2022, Lu2024}, and both are influenced by the enjoyment of effortful cognitive activities (\ncs{})~\cite{West2008, Xiao2021, Wu2023}. To examine if \sys{} was potentially more helpful for the more vulnerable demographic, we conducted correlation analyses of both behavioral and attitudinal measures with these attributes. Pearson's Correlation ($r$)~\cite{Pearson1896} was used in all cases except for the \ed{}, which had a non-normal distribution. In this case, Spearman’s Rank Correlation ($r_s$)~\cite{Spearman1904} was used instead.

\paragraph{Interactions in the \sys{} Task}

All the interactions in the \sys{} condition (see Table~\ref{tab:interactions}) was aggregated for each participant as a behavioral measure. A significant negative correlation was found with the \ncs{} (\Pearson{16}{-.593}{=.010}), but not with the \ed{} (\Spearman{16}{.137}{=.589}) and \ia{} (\Pearson{16}{-.260}{=.298}).

\paragraph{Likability of the \sys{} Features}

The like ratings of all the features (see Section~\ref{sec:postsurvey}) was averaged as the first attitudinal measure. A significant positive correlation was found with \ia{} (\Pearson{16}{.678}{=.002}), but not with the \ed{} (\Spearman{16}{.231}{=.356}) and \ncs{} (\Pearson{16}{-.206}{=.411}).

\paragraph{Usefulness of the \sys{} Features}

The usefulness ratings of all the features (see Section~\ref{sec:postsurvey}) was averaged as the second attitudinal measure. A significant positive correlation was found with \ia{} (\Pearson{16}{.700}{=.001}), but not with the \ed{} (\Spearman{16}{.144}{=.568}) and \ncs{} (\Pearson{16}{-.099}{=.695}).

\subsection{Interview}
\label{sec:interview}

We draw insights from the interview to extend the quantitative results from the tasks and post-survey.
Affinity diagramming~\cite{SCUPIN1997, Krause2024} was used to group the responses into thematic clusters that we detail with quotes from the participants.

\subsubsection{\sys{} Enhances Critical Thinking and Confidence in Writing, With Knowledgeable Users Being Less Dependent}

Most participants reported finding it easier to complete the task with \sys{} (\pcount{13}). We found that \sys{} generally encouraged critical thinking by provoking participants to delve more deeply into the content. Participants mentioned that \sys{} \pquote{makes analysis more systematic}{5}, and that it helps them \pquote{think [in a] more logical and unbiased, \ldots{} rational, more generalizing way}{8}.
For participants that found their written messages for the \sys{} task more satisfactory (\pcount{10}), having improved information search outcomes with \sys{} made them feel more assured about the opinions they expressed in their messages. The probe feature helped them to \pquote{know where to look into, \ldots{} [and affirm] what I have in my mind}{3}. Seeing questions in the chat feature that corroborated their own also \pquote{reinforced that I was going in the right direction}{5}.
More knowledgeable participants, however, expressed a lower dependence on \sys{} in composing their messages. Few participants reported to have knowledge on both topics (\pcount{2}), while most were more well-versed in one than the other (\pcount{11}). Being knowledgeable helped them to \pquote{know what and how to argue}{12} such that they \pquote{didn't feel the need to use the AI}{3}. On the other hand, a lack of knowledge caused conscientious users to feel \pquote{less confident about commenting}{5}. They also mentioned how it affects their use of \sys{} as it \pquote{took me a long time to search and think of a direction and find the right keyword [for the probe]}{9}.

\subsubsection{\sys{} Assists in Evaluating Content, but Inaccuracies and Usability Issues Hinder Use}
\label{sec:Changestothefeatures}

Participants generally appreciated \sys{} as an auxiliary tool that revealed their oversights, affirmed their doubts, and aided in resolving them.
However, they also encountered inaccurate, irrelevant, and repeated content that diminished the usability of \sys{}.
The sentiments of participants' responses for each feature were clustered into positive, negative, mixed (both positive and negative) and unclear (ambiguous sentiment).

\paragraph{Highlight Feature}

Sentiments toward the highlight feature were split (Positive: 3, Negative: 4, Mixed: 10, Unclear: 1).
Participants liked that the highlight feature helped distinguish questionable sentences that could be targeted in their arguments. The color coding and formatting differentiates fallacies spotted by the AI and those raised by users which was appreciated \pquote{cause someone can also fact check the AI or bring up a point they missed}{5}.
Those who disliked the highlight feature instead found it messy and confusing. Errors were noted with \hai{} where \pquote{some unsubstantiated statements [] were not highlighted}{1}. There was concern regarding \huser{} where \pquote{If everyone can highlight \ldots{} How would it scale?}{5}. The \hsays{} were also thought to be verbose, coming across as either \pquote{too technical}{6} or unhelpful as \pquote{a lot of [it] is saying the same thing [which] \ldots{} is quite obvious}{2}.
To improve the feature, participants suggested \pquote{having a toggle to turn on or off the highlights, cause I don't want to immediately know the fallacies}{5}, to \pquote{maybe [have] lesser of everything in a paragraph \ldots{} summarize [the highlights] in one button}{8}, and to be able to contest the fallacy whereby if \pquote{someone suggests \squote{appeal to emotions}, and if sufficient users agree, [then] change it permanently}{3}.

\paragraph{Probe Feature}

Sentiments toward the probe feature were more positive (Positive: 7, Negative: 1, Mixed: 9, Unclear: 1) with nearly everyone having something good to say about it.
Participants favoring the feature cited the convenience of conducting searches as they \pquote{don't have to switch browsers \ldots{} [which can] be disruptive}{3}. The \psug{} aligned with what they were thinking of and the results from \pwrite{} were at times even better than their own questions. A participant expressed that these made it \pquote{super easy to contextualize what the author might be trying to say}{5}.
Those with poorer perceptions of the feature pointed out certain shortcomings. \pwrite{} was perceived to be \pquote{not really useful cause it was just a paraphrase [of what they entered]}{11}. The results from \psug{} \pquote{felt too broad or repetitive \ldots{} [and] did not tie deeply into the complexity of the topic}{14}. For the \pweb{}, \pquote{the references are not very reliable [as they are] not journal articles}{12}.
Additional functionalities were desired such as to \pquote{widen the questions and add more facts, data and figures}{10}.

\paragraph{Chat Feature}

Sentiments towards the chat feature were also rather positive (Positive: 7, Negative: 2, Mixed: 9, Unclear: 0).
Participants who liked the chat feature predominantly mentioned that it guides their thinking. The \cfood{} \pquote{adds thinking points on how to approach my stance and \ldots{} gives [me] ways to analyze the highlighted content}{5}. It also helps them to \pquote{see things from different angles that I have not thought of}{12}. For the \cdis{}, having the chats \pquote{localized makes more sense so that people know what they are referring to}{13}.
As for the negative feedback, the \cfood{} could at times be too generic such that they \pquote{lack specificity or nuance to the arguments being discussed}{14}.
The value of the \cdis{} was questioned as many platforms already have their own commenting sections. Participants mentioned that \pquote{There is already too much to read.}{2} and that it \pquote{might get messy}{8}. Another concern was that \pquote{commenting on one sentence doesn't give the full context}{15}.
Participants remarked that the chat feature could share similar functionalities with the probe feature, suggesting that the \pquote{\cfood{} can \ldots{} become a search term for [\pwrite{}]}{1}, and that it would be \pquote{good to pre-populate [\cfood{} with answers like the \pweb{}]}{11}.

\subsubsection{\sys{} Can Be Censorious, Misaligned, and Inhibitive}

While \sys{} was generally perceived favorably, a few participants further expressed specific frustrations and concerns.
There was aversion towards the use of \sys{} in online forums as it was thought to obstruct the freedom of expression. A participant stated that \pquote{The author is writing their own opinion. [This] doesn't mean it's correct or wrong. Why do I need [\sys{}] to say \squote{I like orange shoes?}\hspace{0mm}}{2}. They perceived online forums as spaces of free speech and felt that \sys{} placed certain impositions by calling out fallacious content.
Frustration was expressed by another participant with how \sys{} aimed for wider and deeper consideration of content whereas they sought for simplification. They said, \pquote{I feel like it's not for me cause it's [giving me] more and more questions whereas I keep looking for more and more shortcuts and answers \ldots{} I like a simple answer like in Occam's razor \ldots{} I don't like expansion and having to explain the problem. Just freaking solve it.}{4}. This shows a mismatch between \sys{}'s objectives and the user's goals.
A concern was also raised that \sys{} might stifle independent thinking. A participant noted that \sys{} segregates the texts such that \pquote{it gets you distracted. People would just focus on the highlighted portion and stop thinking for themselves. [There] could be a case where they think just the highlight is the iffy part and everything else is accurate.}{3}. With this, \sys{} would have been counterproductive.

\subsubsection{\sys{} May Be Better Suited for Contexts That Demand High Precision}

When asked about making \sys{} publicly available, many participants expressed support (\pcount{7}). In forums where a variety of opinions collide, it was mentioned that \sys{} can \pquote{help people to be more mindful, think more, explore more, \ldots{} and not be complacent with what they are reading}{1}. 
A few participants instead said that they would not use it (\pcount{3}), with one mentioning that they \pquote{read online forums to see the range of diverse and extreme opinions}{6}, and another stating that \pquote{there is a lot of anti-AI sentiment}{3}.
For the others, they commented that it depends on the context of use (\pcount{8}). They felt that \sys{} would not be useful for \pquote{topics on stupid random troll stuff}{8} and \pquote{maybe not everyday [topics], maybe politics or economics}{11}. Some participants remarked that they would use it in alternative scenarios like to \pquote{analyze journal articles}{12}, when they are \pquote{googling for news}{7}, or browsing \pquote{social media news pages [with] so-called local news sites \ldots{} [where] the information may not be 100\% correct}{2}. A participant also suggested that \pquote{in terms of teaching like in university courses, [\sys{}] might be good for class discussions}{18}.

\subsubsection{Trust Towards \sys{} Was Generally Reserved, While Accuracy Was Perceived to Be Adequate}
\label{sec:generalperceptions}

To calibrate users' expectations of the performance of \sys{}, we provided its accuracy in detecting fallacies (see Figure~\ref{fig:extension3}). When we asked participants about how this influenced their views of \sys{}, only one participant remarked that this made them \pquote{have more trust}{5}.
The others expressed various levels of skepticism (\pcount{13}), either because they were a naturally cynical person, knew that AI could not be 100\% correct, or wanted to judge the accuracy only after using it. A few participants were not particularly concerned about accuracy at all (\pcount{4}).
When asked about their actual experience from using \sys{}, some participants thought it was \pquote{pretty accurate}{9} and \pquote{maybe 70-80\% generally}{10}. As \sys{} supports the user to come to an independent conclusion of the content, participants expressed tolerance towards mistakes it made as they could make up for oversights. Despite noting errors in the highlight feature, a participant remarked that \pquote{It's not intended to replace critical thinking, but highlight potential issues. So even if it's not perfect, it focuses on areas for scrutiny that users can check.}{14}. Fewer participants remarked on the probe and chat features, finding them acceptable. One, in particular, felt that \pquote{The accuracy of the probe and \cfood{} is dependent on the highlight.}{4}.

\subsubsection{Fallacies Were Easily Picked Up While Using \sys{}, and the Fallacy Fallacy Was Largely Avoided}
\label{sec:fallacyfallacy}

Most participants had no issue understanding the fallacies (\pcount{11}).
Several participants further expressed that they became familiar with the fallacies with time. As put by a participant, \pquote{I keep reading the same fallacy so it's quite repetitive. At the end of the day, I don't need [\sys{}]. Over time, I can pinpoint certain statements by myself.}{2}.
With \sys{} using fallacies to avoid placing veracity judgments and to assist people in evaluating content, we asked participants whether they had fallen into the Fallacy fallacy, where content with fallacies is perceived as false, even when it is not (see Figure~\ref{fig:extension3}).
The majority of participants did not experience the Fallacy fallacy for reasons such as having \pquote{prior knowledge}{5}, being \pquote{more critical}{4}, perceiving the fallacies to simply indicate that \pquote{the argument is not making [any] sense}{12}, and having the view that \pquote{fallacies is not about being false [but] about being incomplete. It can be partially true.}{18}. A few participants (\pcount{3}), however, reflected that they had committed the fallacy. One of them said, \pquote{I would be susceptible to it cause it's easy to dismiss something as false if it contains a fallacy. I'm sure I did.}{6}.

\section{Discussion}

We consolidate insights from the results to discuss how \sys{} supports critical thinking, the potential backfiring effects of \sys{} and their mitigations, and possible avenues of development for \sys{}.

\subsection{Unpacking the Effectiveness of Iffy-Or-Not}

\subsubsection{\sys{} Supports Critical Thinking}
During the interview, several participants remarked on how they were more critical of the content they read while using \sys{}.
The highlights raised their attention to questionable parts of the texts which they may have glossed over otherwise.
Having the convenience of the probe situated beside the highlight made it easier for them to look up information. The alternative of opening new tabs to conduct searches would have been disruptive given that context switching drives cognitive overhead when attention is split across multiple contexts~\cite{LEROY2009168}. Having the \pwrite{} search embedded in the page also makes it easy for them to make revisions to their queries. Situated searches have been found to improve efficiency in generating the desired query~\cite{Aula2005}.
Apart from convenience, participants found that \sys{} expanded the way they considered the highlighted content. The \psug{} provided more angles of inquiry, at times mirroring the thoughts they already had. The \cfood{} questions offered alternative perspectives that diverged from their personal views.

However, participants also had accuracy issues with \sys{} such as fallacious content that should have been highlighted but were not. Some also felt that the queries and questions in the probe and chat features were at times out of context. These made them question the performance of \sys{} and gradually disengage from the use of certain features that they felt were subpar.

Taken together, \sys{} raises the level of scrutiny towards fallacious content, provides assistance to look for the desired information to fulfil gaps in their knowledge to make an informed evaluation of the content, and to promote deeper and broader thinking about the issue of the content. The strong reasoning and creative capabilities of the LLM promotes the latter two in a serendipitous fashion as it presents valued suggestions to look up or consider the content in ways that participants may not immediately come to or even think of. This is, however, on the basis that \sys{} is performant and up to users' expectations.

\subsubsection{Varying Helpfulness of \sys{}}

We examined whether \sys{} was more helpful for certain users, particularly those who might be less critical of content or more susceptible to misinformation (see Section~\ref{sec:helpfulness}). Generally, \sys{} appears more helpful for users with a lower preference for effortful cognitive activities, greater knowledge of the topic, and stronger media literacy.

The behavioral (\ix{} with \sys{}) and attitudinal (\lik{} and \use{} of \sys{}) attributes were negatively correlated with the \ncs{}, with \ix{} being significantly so, suggesting that \sys{} is more helpful for those with a lower proclivity for effortful thinking. As \sys{} provides thought-provoking questions and eases the information search process, users with a lower \ncs{} could have felt that it was of assistance to them. In contrast, users with a higher \ncs{} could have been more independent thinkers and have accustomed ways of processing information~\cite{Fortier2014}, hence preferring not to be guided, and thereby restricted, in \sys{}'s ways.

Generally, \sys{} seemed slightly more helpful for those with more knowledge on the topics given the weak positive correlations of all the attributes on the \ed{}. As \sys{} can be information-heavy, having more knowledge on the topic could have enabled participants to navigate the content presented more easily as they already had a contextual background to better process and understand them~\cite{Anderson1996}.

An inconsistency was observed for \ia{} which was negatively correlated with \ix{}, but had significant positive correlations with \lik{} and \use{}, suggesting that participants with higher media literacy skills used \sys{} less, yet perceived it more favorably. A plausible explanation for this could be that these users were more efficient in using \sys{} to seek their desired information, and were satisfied with the content \sys{} provides.

Putting these together, \sys{} helps to reduce cognitive load in processing and evaluating information. However, it's usability is impacted by users' level of knowledge on the topic and their skills in assessing and verifying online content.

\subsubsection{Potential Backfiring Effects With \sys{}}

While \sys{} can bolster the critical evaluation of information, the results have also surfaced certain caveats with it that may potentially backfire. Participants have raised views on the features making the forum \dquote{messy} where the highlights can also be distracting. As each feature of \sys{} adds more content to the screen, reading can become an overwhelming experience in platforms like online forums that are already filled with texts. The additional cognitive burden involved to process the added content and engage in information evaluation practices may lead to \textit{information overload}~\cite{Arnold2023}, potentially resulting in the reduced use of \sys{}.

Another issue raised by a participant is that users may be more prone to relying on \sys{} to do the thinking rather than on themselves. Not only does this lie in direct contradiction with the aim of \sys{} by potentially compromising critical thinking, but it further brings in a general problem in human-AI collaboration of \textit{overreliance}. AI has been found to impact decision-making where advice by the AI is followed even when it contradicts available contextual information and the users' own assessment~\cite{Klingbeil2024}. This has also been observed in AI fact-checking tools where inaccuracies in the AI translate to impaired human decision-making on the veracity of the content~\cite{Nguyen2018, Lu2022}. Ensuring appropriate reliance would thus be a challenge for \sys{}.

A participant also mentioned how the highlights may make people assume that non-highlighted content is fine instead. The issue of warnings unintentionally impacting content without them is known as the \textit{implied truth effect} where unlabeled content is perceived to be more accurate than it really is~\cite{Pennycook2020a}. While this effect centers on the veracity of the content while the highlights by \sys{} are on fallacies which are fundamentally unconcerned with veracity, as discussed in Section~\ref{sec:fallacyfallacy} on the Fallacy fallacy, it is not unusual for people to subconsciously make that association. It is thus understandable that users may interpret the lack of highlights as the absence of problems, thereby overlooking them.

\subsection{Mitigations for the Backfiring Effects of Iffy-Or-Not}

\subsubsection{Information Overload}
While the issues raised above are nontrivial, perhaps the most addressable concern is that of information overload--to reduce the amount of information that \sys{} presents. During the interview, participants tended to refer to the probe and chat features together, and some further commented that one can incorporate features from the other. These suggest that there are parallels between the features which might merit combining them into one. We envision removing the \cdis{} from the chat feature, as it has the least impact given that social platforms have their own, and integrating the \cfood{} into the probe feature, which could use an accordion design to optimize screen estate. Borrowing a participant's suggestion from Section~\ref{sec:Changestothefeatures}, another option is to have toggles for each feature to give users the agency to shape what they see from \sys{} as best fits their current purpose.

\subsubsection{Implied Truth Effect}
Regarding the implied truth effect, we borrow another idea that involves reducing the highlights. While \sys{} already has such a feature with the \hchart{}, it can be made smaller and shifted to the end of the text instead where interacting with it would then bring up the highlights. By adding another layer of abstraction to surface the highlights rather than presenting them on the outset, we seek to lessen the visual impact of the highlights that might lead to biased cognitive processing. Furthermore, the workflow of \sys{} is modified by encouraging users to read and engage with the content first before considering the fallacies in them. While this change might alleviate the implied truth effect, further testing is required as people have also been shown to be resistant to correction after forming their initial beliefs~\cite{Ecker2022} and this change in workflow which prioritizes unguided reading where people would form their views of the text first could manifest undesirably in another way.

\subsubsection{Overreliance}
For the study, we sought to mediate people's trust in \sys{} by being transparent about its fallacy detection accuracy (Figure~\ref{fig:extension3}). As seen in Section~\ref{sec:generalperceptions}, participants expressed varying levels of skepticism to \sys{}. This signals that they did not fully trust \sys{} which suggests that overreliance could be less of an issue. Nevertheless, it remains a crucial point to look at. Prior work has suggested that \dquote{slow} algorithms that impose a waiting time between their activation and response can elicit analytical thinking which mitigates overreliance on the AI~\cite{Park2019, Rastogi2022}. Mounting on the adjustment to the highlights mentioned above, we further suggest adding a delay in showing the highlights after users have interacted with the \hchart{} to trigger them. While this may seem like a simple adjustment, it is noteworthy that people find such algorithms to be unfavorable~\cite{Bucinca2021} which suggests a need to carefully calibrate the response time for better human alignment.
Borrowing another idea, the highlight feature could include a community rating module that allows users to contest the highlighted phrase and the detected fallacy. If enough ratings are accumulated, what was surfaced by \sys{} would be replaced by the majority consensus, with an indicator of such a change. This aligns with contestable AI that leverages human intervention to attenuate incorrect decisions by the AI~\cite{Alfrink2023}.

\subsection{Avenues of Development for Iffy-Or-Not}

\subsubsection{Misalignments of \sys{} With Users}

From the results, a few participants expressed frustration with \sys{} as it's mechanism of promoting deeper thinking ran counter to how they think. One participant mentioned being a follower of Occam's Razor where they \pquote{want to simplify things, not expand them, [and] want answers not questions}{4}. They likened the \sys{} task to being an argumentative essay assignment. While this is just one participant, it is not unexpected for there to be people with analytical and reasoning preferences that differ from \sys{} that pushes for deeper thought.

Some participants also questioned having \sys{} in the context of online forums. They perceived these to be spaces for opinion expression--even those that are extreme, as well as for entertainment, which do not call for a high level of critical thinking and deep analysis. While \sys{} highlights fallacious content, as discussed earlier, such highlights may be associated with labels of veracity. Such labels have been found to be perceived as censorious and contrary to the purposes of these spaces for the freedom of expression~\cite{Saltz2021}, and participants may have also held similar perceptions toward the highlights. This signals a discrepancy between the actual purpose of \sys{} which is to surface fallacious content to bring them to the user's attention and the perceived purpose of \sys{} by users whereby it judges the highlighted content to be problematic.

Together, misalignments in the modes of thinking and usage contexts are at least two factors that would impact the intention to use \sys{}. This presents opportunities for other AI tools that can cater to different sets of user profiles and information needs.

\subsubsection{Use of \sys{} in Other Contexts}

Several participants reported that they would want to use \sys{} for more formal contexts like reading news articles and academic papers. In these contexts, readers would likely gravitate towards thinking more critically which may be why they associated \sys{} with such purposes. In some ways, \sys{} could perhaps be more valuable in these contexts. Mainstream news websites and academic publications seem to be viewed as trustworthy sources by our participants in how they usually prioritized evidence from them during the tasks. While trust in sources is a heuristic that helps people to quickly filter for credible content~\cite{Metzger2013}, it may also cause them to be less discriminate and critical of the information~\cite{Pornpitakpan2004, Sundar2008}. Despite the presence of journalistic reporting standards and peer review for quality assurance, there is a chance for biased reporting and mistakes to be made which may pass by unnoticed for a reader who trusts the source and thereby extends that trust to the content as well. Having a second eye through \sys{} can help in circumventing such inattentiveness.

\subsubsection{Education and Misinformation Inoculation}

Many participants mentioned that the fallacies surfaced by \sys{} were easily picked up with one of them saying that, with time, they could do it on their own. This alludes to an educational aspect of \sys{}. As it supports a small number of fallacies with the definitions of them in the \hchart{} and the \hsays{} elaborating why the highlighted content was fallacious, participants would have been repeatedly exposed to various examples of the same fallacies and implicitly learned from it, known as the mere exposure effect~\cite{Gordon1983}. Such learning can be long-lasting when they become tacit knowledge~\cite{Reber1989}, and make users more attuned and sensitive to fallacious content even without \sys{}, which we perceive to be the most desirable outcome.

Being more sensitive to fallacious content in texts may further lend towards inoculating them on harmful content like misinformation. Inoculation is the exposure to a weak and controlled dose of something malicious to confer resistance against the real thing~\cite{McGuire1964}. To address misinformation, inoculation by exposing participants to various manipulation techniques have been found to be effective~\cite{Roozenbeek2022}. In that study, several of the manipulation techniques assessed were also fallacies, with the results that the interventions \dquote{improve manipulation technique recognition [and] boost confidence in spotting these techniques}. The parallels that our study has with it suggests that \sys{} may help confer resistance through similar mechanisms.

\subsubsection{Advances for \sys{}}

Beyond the current features, there are several directions in which \sys{} can be extended.

\paragraph{Visualizations} Statistical data is a common way to evaluate and substantiate claims~\cite{Hoeken2009}. While \sys{} at times offers statistical data if they are drawn up in the \pweb{}, some participants expressed a preference towards having figures instead. LLMs have been used to gather and visualize data from the web to better contextualize unsubstantiated statistical claims~\cite{Kim2024}. Such an integration with \sys{} could greatly enhance the information evaluation experience.

\paragraph{Multimodality} Social platforms are often multimedia and the images and videos that proliferate there can be more persuasive~\cite{Shen2015} and credible~\cite{Hameleers2020}, and thus more harmful than texts when their content is problematic. This raises the potential of using large multimodal models that are gaining strong traction in research~\cite{Yin2024}, to translate the functionalities of \sys{} to support these modalities as well.

\paragraph{Critical Writing} Beyond reading, online deliberation also calls for being critical in writing content. Lightweight interface nudges have been shown to facilitate deeper deliberation~\cite{Menon2020} and reflection~\cite{Yeo2024} to enhance the introspection and quality of arguments in the comments. Such nudges could work in tandem with \sys{} when implemented in social platforms to provide a holistic end-to-end experience in engaging critically with online content.

\section{Limitations}

To adjust expectations for \sys{}, only the five most common fallacies in online discussions were detected to assess and share the LLM's accuracy with participants.
However, some fallacies, like against the person or appeal to emotion, tend to be overt and easy for people to catch, thus being less valuable to surface. They are also more prevalent, which can dilute the impact of other highlights. 
As the fallacies that \sys{} raises affect both its experience and value, prioritizing subtler fallacies, such as questionable cause, can be more helpful.

As the study sessions were conducted individually, participants could only chat in the \cdis{} asynchronously, which limited the level of engagement they had. A future work could be to hold small group studies to assess the collaborative interactions and dynamics as several people use \sys{} at the same time.
A related issue is that we did not design mechanisms to catch fallacies that may arise as users converse in the \cdis{}. A future development of \sys{} could be to actively review chat messages for fallacies as they are written, and to prompt users to revise them before posting~\cite{Zhang2023}.

The study used Reddit r/changemyview threads that can be highly opinionated, and a hotbed for fallacies.
We chose Reddit for its popularity, and the deliberative nature of r/changemyview~\cite{Tan2016}, to provide a more natural setting for our participants who were familiar with the platform. However, the results may thus not generalize well to other contexts.
As we conducted a user study to gain a deeper understanding of \sys{}'s effects, the findings are also limited by our task design and sample size. For future work, an in-the-wild deployment with a large sample will be valuable to understand how different users interact with \sys{}, and what they use it for.

\section{Conclusion}

Social platforms have become spaces for deliberation and opinion formation but the quality of content on them is compromised by misinformation. To enhance the evaluation of online information, we present Iffy-Or-Not (\sys{}), a browser extension that supports critical reading and evaluation of online content by surfacing fallacies in the texts (highlight feature), supporting users to explore further by enhancing the search initiation and query generation process (probe feature), and offering insightful and diverse questions to provoke deeper thinking (chat feature).

A technical evaluation of the LLM powering \sys{} shows that it is able to detect five common fallacies in online discussions with an average 84\% accuracy.
Through a user study (\N{18}), we found that highlighting fallacious content draws users’ attention towards the highlights and makes them more cynical towards them. Having search probes situated in the forum enhances the convenience of initiating search and some users expressed appreciation of the search queries suggested by \sys{} as they aligned with what they had in mind, and were at times preferable to their own. The questions posed in the chat also encouraged them to think further about the fallacious content and from more diverse angles. Overall, \sys{} was able to enhance users’ critical evaluation of the information they consume. However, users also found errors by \sys{} which impacted their views and usage of them. Misalignment between the goals of \sys{} and their personal information goals and thinking dispositions also made them question the value of the extension. Potential backfiring effects were raised on information overload, the implied truth effect and overreliance and we discuss potential ways to mitigate them. We also offer various avenues of improvements in \sys{} and beyond it.

As online deliberation and opinion formation grapples with unsubstantiated content and misinformation, AI tools can help to cover the bases by scaling with the ever-increasing amount of information produced online, and supplementing the bounded cognitive processing of humans in evaluating information. \sys{} is just one example, and the development and assessment of more of such tools will lends towards enriching the design space of human-AI collaboration in consuming online content.

\begin{acks}
Our gratitude to Loh Haw Yuh on assisting with the set up and maintenance of the server and the LLM.
\end{acks}

\bibliographystyle{ACM-Reference-Format}
\bibliography{main}

\appendix
\section{LLM Prompts}
\label{sec:prompts}

The prompt format follows the documentation for Meta Llama 3\footnote{\url{https://llama.meta.com/docs/model-cards-and-prompt-formats/meta-llama-3/}}. The same role is set for all of the prompts: \texttt{You are a critical thinker.}

\subsection{Prompt for AI Highlights: Identifying Fallacies}
\label{app:detectprompt}

When the \dquote{Find Iffy Content} button is triggered, fallacies in the text are detected as \hai{}s. This has a temperature setting of \texttt{0}, and a maximum number of \texttt{512} new tokens.

\begin{lstlisting}
<|begin_of_text|><|start_header_id|>system<|end_header_id|>You are a critical thinker.<|eot_id|><|start_header_id|>user<|end_header_id|>There are five fallacies with their definitions and examples below.

{
  "fallacy": "questionable cause",
  "definition": "Concluding that one thing caused another simply because they are regularly associated.",
  "examples": ["Children who play violent video games act more violently than those who don't.", "President Kumail raised taxes, and then the rate of violent crime went up. Kumail is responsible for the rise in crime.", "People who eat yogurt have healthy guts. If I eat yogurt I will never get sick."]
},
{
  "fallacy": "ad populum",
  "definition": "Affirming that something is real or better because the majority in general or of a particular group thinks so.",
  "examples": ["Everyone seems to support the changes in the vacation policy, and if everyone likes them, they must be good.", "Since 88% of people polled believe in UFOs, they must exist.", "Don't be the only one not wearing Nike!"]
},
{
  "fallacy": "ad hominem",
  "definition": "Attacking the person or some aspect of the person making the argument instead of addressing the argument directly.",
  "examples": ["Only a selfish, non-caring person would believe that this is ok.", "You can't believe Jack when he says there is a God because he doesn't even have a job.", "The reason our company never makes any money is because we have a buffoon running it!"]
},
{
  "fallacy": "appeal to emotion",
  "definition": "Manipulating the reader's emotions in order to win the argument in place of a valid reason.",
  "examples": ["Regime vs. government Pro-death vs. pro-choice", "It is an outrage that the school wants to remove the vending machines. This is taking our freedom away!", "Television Advertisement: Get all of your stains out by using new and improved Ultra Suds and wash your blues away!"]
},
{
  "fallacy": "appeal to authority",
  "definition": "Using an alleged authority who is not really an authority on the facts relevant to the argument as evidence.",
  "examples": ["My professor, who has a Ph.D. in Astronomy, once told me that ghosts are real. Therefore, ghosts are real.", "There is definitely a link between dementia and drinking energy drinks. I read about it on Wikipedia.", "Coke is not as healthy for you as Pepsi. Besides, Britney Spears drinks Pepsi, so it must be healthier than Coke."]
}

For the text below, check if it contains a fallacy. If there is no fallacy, return only one word: nothing. Else, return each fallacy using the template:

{
  "part": "quote verbatim the part of the text with the fallacy",
  "fallacy": "the fallacy",
  "explain_short": "explain the fallacy concisely in up to 30 words"
  "explain_long": "explain the fallacy in detail in up to 70 words"
},

The text:

--The text goes here--<|eot_id|>
<|start_header_id|>assistant<|end_header_id|>
\end{lstlisting}

\subsection{Prompt for AI Highlights: Suggested Queries and Food For Thought}
\label{app:suggestedqueries}

For each \hai{}, the \psug{} and \cfood{} questions are generated. This has a temperature setting of \texttt{0.7}, and a maximum number of \texttt{512} new tokens.

\begin{lstlisting}
<|begin_of_text|><|start_header_id|>system<|end_header_id|>You are a critical thinker.<|eot_id|><|start_header_id|>user<|end_header_id|>A part of the text contains a fallacy.

{
  "text": "--The text goes here--",
  "part": "--The part goes here--",
  "fallacy": "--The fallacy goes here--"
}

For each part, do the following.

1) critical_questions: Create a list of 8 distinct questions to critically consider, evaluate and verify the part of the text with the fallacy.
2) critical_queries: Create a list of 3 distinct search queries, written in complete sentences, using the part of the text with the fallacy that can be entered into a search engine to critically consider, evaluate and verify it.

Return the results using the template.

{
  "critical_questions":  [question, question, ...],
  "critical_queries": [query, query, ...]
}<|eot_id|>
<|start_header_id|>assistant<|end_header_id|>
\end{lstlisting}

\subsection{Prompt for Write Own Query}
\label{app:revisedqueries}

When a user enters their search terms to \pwrite{}, the revised queries are generated. This has a temperature setting of \texttt{0.7}, and a maximum number of \texttt{256} new tokens.

\begin{lstlisting}
<|begin_of_text|><|start_header_id|>system<|end_header_id|>You are a critical thinker.<|eot_id|><|start_header_id|>user<|end_header_id|>Some search terms have been entered for a part of the text.

{
  "text": "--The text goes here--",
  "part": "--The text goes here--",
  "search_terms": "--The search terms go here--",
}

Using the search terms, create a list of 3 distinct and concise search queries for the part of the text that can be entered into a search engine to critically consider, evaluate and verify it.

Return the results using the template.

{
  "revised_queries":  [query, query, ...],
}<|eot_id|>
<|start_header_id|>assistant<|end_header_id|>
\end{lstlisting}

\subsection{Prompt for Suggested Queries: Extracting Web Content}
\label{app:extractweb}

When one of the \psug{} is triggered, to generate the \pweb{}, the first part involves extracting the most relevant web content. This has a temperature setting of \texttt{0.7}, and a maximum number of \texttt{512} new tokens. Web content beyond 2500 words is cut due to the 4096 maximum tokens limit of the LLM.

\begin{lstlisting}
<|begin_of_text|><|start_header_id|>system<|end_header_id|>You are a critical thinker.<|eot_id|><|start_header_id|>user<|end_header_id|>There is a search query for a text.

{
  "text": "--The text goes here--",
  "search_query": "--The search query goes here--",
}

Based on the search query, extract verbatim key part(s) of the text. Each part should be one to three sentences long. Extract at least one part and at most five parts from the text. Be highly selective and extract only parts that are the most relevant to the search query. Prioritize supporting evidence and examples when available.

Return the results using the template.

{
  "extracts":  [extract, ...],
}<|eot_id|>
<|start_header_id|>assistant<|end_header_id|>
\end{lstlisting}

\subsection{Prompt for Suggested Queries: Summarizing Web Content}
\label{app:summarizeweb}

When one of the \psug{} is triggered, to generate the \pweb{}, the second part involves summarizing the extracted web content. This has a temperature setting of \texttt{0.7}, and a maximum number of \texttt{256} new tokens.

\begin{lstlisting}
<|begin_of_text|><|start_header_id|>system<|end_header_id|>You are a critical thinker.<|eot_id|><|start_header_id|>user<|end_header_id|>There are three text extracts for a search query.

{
  "extracts": {
    1: --The first extract list goes here--,
    2: --The second extract list goes here--,
    3: --The third extract list goes here--,
    },
  "search_query": "--The search query goes here--",
}

Based on the search query, summarize the text extracts. The summary should be at least 80 words and at most 150 words. Be highly selective and summarize parts that are the most relevant to the search query. Prioritize supporting evidence and examples when available.

Return the results using the template.

{
  "summary": "the summary",
}<|eot_id|>
<|start_header_id|>assistant<|end_header_id|>
\end{lstlisting}

\subsection{Prompt for User Highlights: Suggested Queries and Food For Thought}
\label{app:foodforthought}

When a user marks their own iffy content as a \huser{}, the \psug{} and \cfood{} questions are generated. This has a temperature setting of \texttt{0.7}, and a maximum number of \texttt{512} new tokens.

\begin{lstlisting}
<|begin_of_text|><|start_header_id|>system<|end_header_id|>You are a critical thinker.<|eot_id|><|start_header_id|>user<|end_header_id|>A part of the text has been marked as suspicious with the reason provided.

{
  "text": "--The text goes here--",
  "part": "--The part goes here--",
  "reason": "--The reason goes here--",
}

For each part, do the following.

1) critical_questions: Create a list of 8 distinct questions to critically consider, evaluate and verify the part of the text based on the reason.
2) critical_queries: Create a list of 3 distinct search queries, written in complete sentences, using the part of the text based on the reason that can be entered into a search engine to critically consider, evaluate and verify it.

Return the results using the template.

{
  "critical_questions":  [question, question, ...],
  "critical_queries": [query, query, ...]
}<|eot_id|>
<|start_header_id|>assistant<|end_header_id|>
\end{lstlisting}

\newpage
\onecolumn
\section{Normalized Confusion Matrices}
\label{app:cmatrix}

\begin{figure*}[!htb]
    \centering
    \includegraphics[width=.75\textwidth]{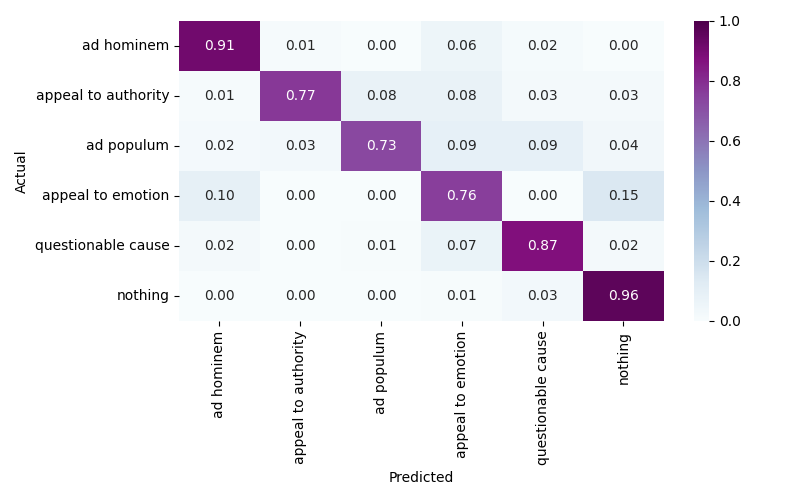}
    \caption{Normalized confusion matrix for the full data (\N{630})}
    \label{fig:ncm_full}
    \Description{A figure on the normalized confusion matrix for the full data where ad hominem had the highlight classification accuracy of 91 percent and ad populum has the lowest at 73 percent.}
\end{figure*}

\begin{figure*}[!htb]
    \centering
    \includegraphics[width=.75\textwidth]{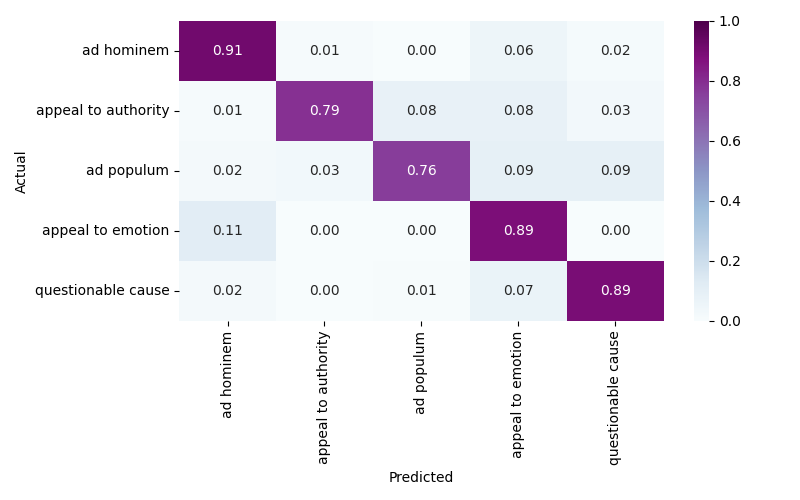}
    \caption{Normalized confusion matrix for the subset data (\N{517})}
    \label{fig:ncm_subset}
    \Description{A figure on the normalized confusion matrix for the subset data where ad hominem had the highlight classification accuracy of 91 percent and ad populum has the lowest at 76 percent.}
\end{figure*}

Normalized confusion matrices of the LLM in identifying logical fallacies for the full data (Figure~\ref{fig:ncm_full}) and the subset data (Figure~\ref{fig:ncm_subset}). Latin versions were used for against the person (ad hominem) and appeal to popularity (ad populum) to corroborate the labels in the LOGIC dataset~\cite{jin2022logical}.

\end{document}